\documentclass[fleqn,12pt,twoside]{article}
\usepackage{espcrc1}

\usepackage{graphicx}
\usepackage{epsfig}
\usepackage[figuresright]{rotating}

\def \la {\mathrel{\vcenter
     {\offinterlineskip \hbox{$<$}\hbox{$\sim$}}}}

\newcommand{\AmS}{{\protect\the\textfont2
  A\kern-.1667em\lower.5ex\hbox{M}\kern-.125emS}}

\title{Neutrino-driven supernovae:\\ An accretion instability in a 
       nuclear physics controlled environment} 

\author{\hbox{$\!\!$
        H.-T.~Janka\address[MPA]{Max-Planck-Institut f\"ur Astrophysik,\\
        Karl-Schwarzschild-Str. 1, D-85741 Garching, Germany}%
        \thanks{This work was supported by the Sonderforschungsbereich
                (SFB) 375 ``Astro-Particle Physics'' and by the
                SFB-Transregio 7 ``Gravitational Wave Astronomy'' of
                the Deutsche Forschungsgemeinschaft.},
        R.~Buras\addressmark,
        F.S.~Kitaura Joyanes\addressmark,
        A.~Marek\addressmark,
        M.~Rampp\addressmark,
        $\,$and~L.~Scheck\addressmark 
        }
        }
       
\begin{document}

\maketitle

\begin{abstract}
New simulations demonstrate that low-mode, 
nonradial hydrodynamic instabilities of the accretion shock 
help starting hot-bubble convection in supernovae and thus
support explosions by the neutrino-heating mechanism.
The prevailing conditions depend on the high-density equation 
of state which governs stellar core collapse, core
bounce, and neutron star formation.
Tests of this sensitivity to nuclear physics variations
are shown for spherically symmetric models. 
Implications of current explosion models for 
r-process nucleosynthesis are addressed.
\end{abstract}

\section{INTRODUCTION}

Supernova (SN) models in spherical symmetry (1D) --- even the latest
ones with ``Boltzmann neutrino ($\nu$) 
transport''~\cite{janka.ref_BTsims} --- do not explode, neither by
the prompt bounce-shock mechanism, nor by the delayed 
neutrino-heating mechanism. Suggestions that neutron-finger
convection in the neutron star (NS) could enhance the $\nu$
luminosity and thus the heating behind the shock~\cite{janka.ref_wilmay}
could not be confirmed by detailed analysis~\cite{janka.ref_bru04}.
Also Ledoux convection in the nascent NS seems to be 
ineffective in this respect~\cite{janka.ref_jan04a,janka.ref_buras03}. 
A reduction of the $\nu$ opacity by several 10\% at densities
below $\sim\,$10$^{13}$g$\,$cm$^{-3}$ would also allow $\nu$'s
to leave the accretion layer more easily and could produce 
explosions~\cite{janka.ref_froeh04}, but a physical effect
of corresponding size is not known.

Multi-dimensional simulations of core-collapse SNe
revealed that hydrodynamic overturn behind the stalled accretion
shock develops on a relevant timescale and increases the efficiency 
of the $\nu$-heating mechanism, enabling explosions
even when spherical models fail~\cite{janka.ref_SN2D}.
Initially there
was hope that this new twist in the theory of SNe
was the long-sought guarantee for ``robust'' explosions.
But the first generation of multi-dimensional models employed
a simplified treatment of the $\nu$ transport --- at best 
by grey flux-limited diffusion schemes --- which fell much
behind the sophistication of the transport in non-exploding 
1D models. It was therefore suspected that the transport
approximations might have contributed to the success
of the multi-dimensional simulations~\cite{janka.ref_mezz98}.

Recently, two-dimensional (2D) models have become available with
a significant improvement of the $\nu$ transport by solving
the energy-dependent equations of $\nu$ number, energy, and
momentum, making use of closure relations obtained from a model
Boltzmann equation~\cite{janka.ref_jan04a,janka.ref_jan04b}. 
The description of
$\nu$-matter interactions has thus reached a new level of
accuracy and refinement, although the transport
is still treated to be essentially radial (lateral flux components
are ignored in the moments equations but terms associated
with lateral velocities and lateral gradients of $\nu$
pressure are included). Simulations with thus improved
treatment of the $\nu$ physics could not find explosions 
despite of convective activity in the $\nu$-heating layer
behind the SN shock~\cite{janka.ref_buras03}, confirming the
suspicion that radical transport approximations might have 
favored the rapid and powerful explosions seen in other 2D 
and 3D models~\cite{janka.ref_SN2D,janka.ref_SN3D}.

So what is missing in the currently most advanced SN
models which fail to produce explosions? In the following, we
shall first briefly summarize the status of core-collapse
modeling by the Garching group. In particular,
we shall discuss that any limitation of the angular wedge of 2D
simulations to less than 180 degrees (e.g. to 90 degrees as in
many previous simulations) imposes artificial constraints
to the fluid flow and suppresses large-scale modes that can
play an important role for the growth of convection. It is 
further demonstrated that even a ``modest'' amount of rotation
in the SN core, in the ballpark of predictions of current 
stellar evolution models~\cite{janka.ref_heger04}, may
have an important impact on the postbounce evolution. Finally, 
we shall elaborate on uncertain aspects of the nuclear physics
which the SN shows sensitivity to.

\begin{figure*}[t!]
\tabcolsep=0.5mm
\begin{tabular}{lr}
$\ \ $
\epsfclipon \epsfig{file=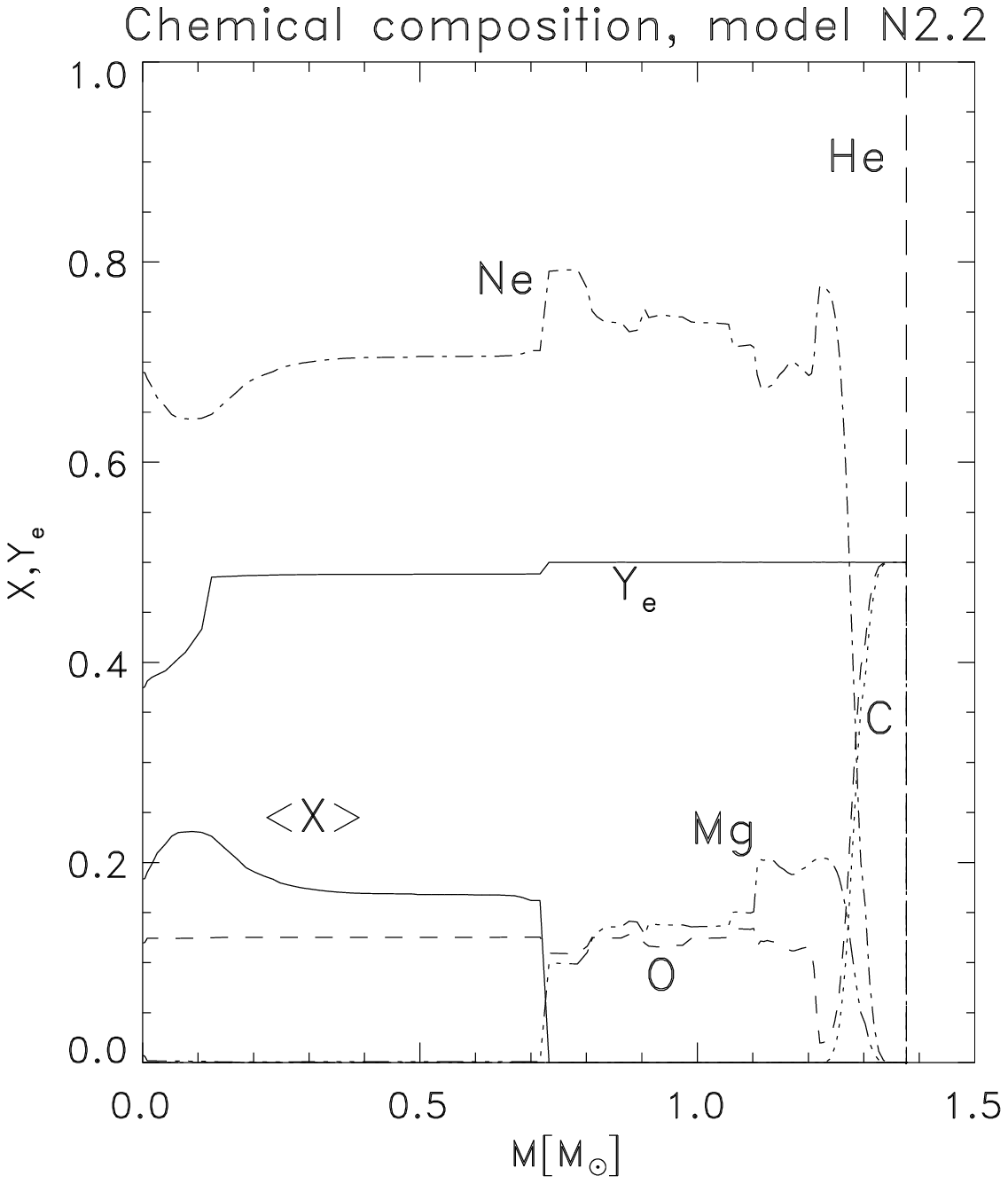,width=0.425\hsize}
$\ \ $
   \raisebox{5.0cm}{
         \parbox[t]{0.45\hsize}{
\epsfclipon \epsfig{file=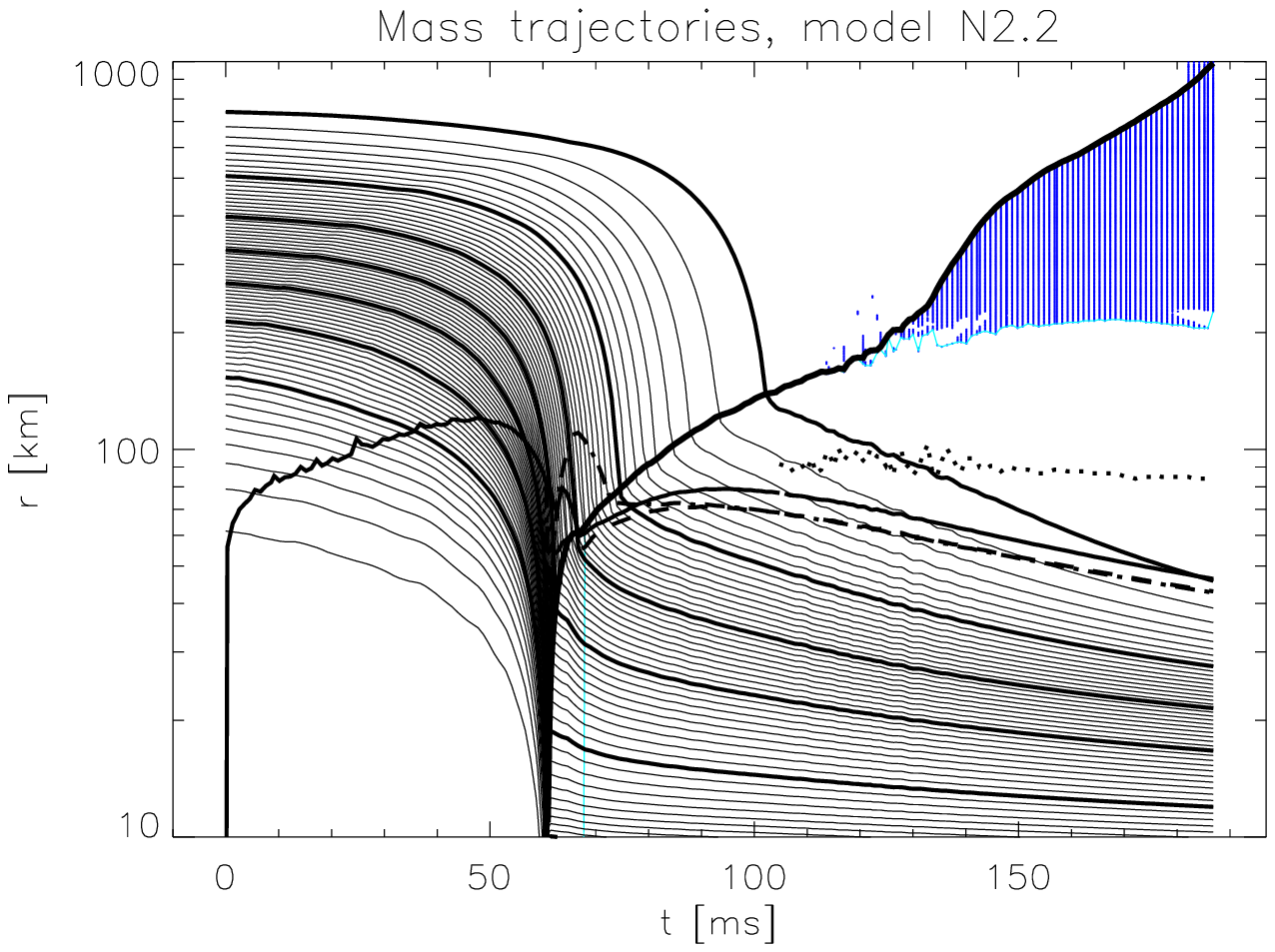,width=\hsize}
\epsfclipon \epsfig{file=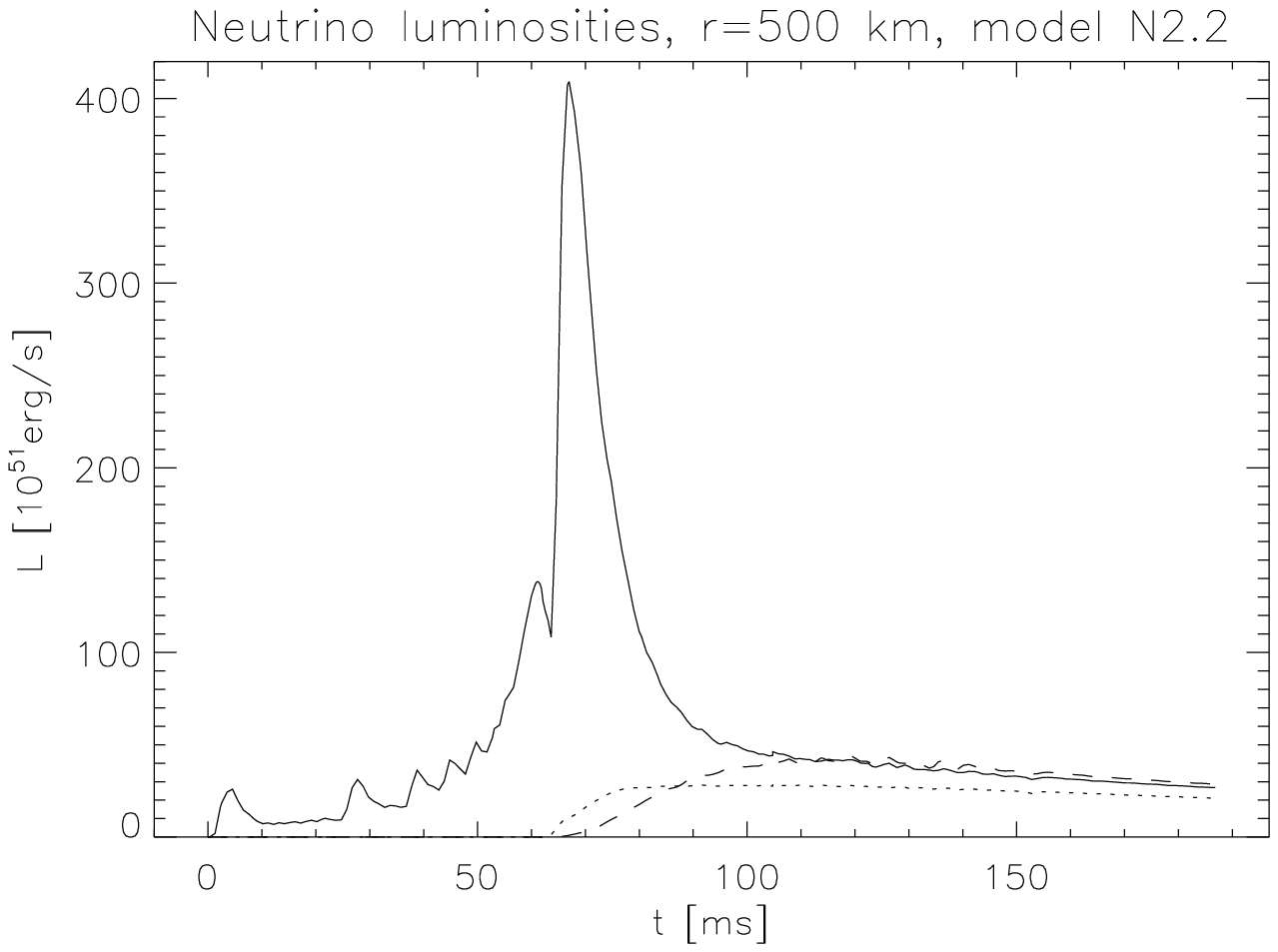,width=\hsize}
}}
\end{tabular}
\vspace{-25pt}
\caption{\small 
{\em Left:} Composition in the highly degenerate 1.38$\,M_{\odot}$
core of an 8--10$\,M_{\odot}$ progenitor~\cite{janka.ref_nom87}
with O, Ne and Mg in the central region enclosed by a C-O shell.
$\left\langle X \right\rangle$ denotes the mass fraction of a
representative neutron-rich heavy nucleus that appears in
nuclear statistical equilibrium, and $Y_e$ is the electron
fraction.
{\em Right, top:} Mass trajectories, shock position (thick solid line
that rises to the upper right corner of the plot, where it reaches
the surface of the C-O shell), neutrinospheres ($\nu_e$: solid;
$\bar\nu_e$: dash-dotted; $\nu_{\mu},\,\bar\nu_{\mu},\,\nu_{\tau},\,
\bar\nu_{\tau}$: dashed), and gain radius (dotted) for the collapsing  
O-Ne-Mg core. The mass trajectories are equidistantly spaced with
intervals of 0.02$\,M_{\odot}$. The outermost line corresponds to
an enclosed mass of 1.36$\,M_{\odot}$ near the outer edge of the carbon
shell. The (blue) hatched region is
characterized by a dominant mass fraction of alpha particles. 
{\em Right, bottom:} Luminosities of $\nu_e$ (solid),
$\bar\nu_e$ (dashed) and heavy-lepton $\nu$'s
(individually; dotted) as functions of time. The
sawtooth pattern during the collapse phase until about 50$\,$ms
is a numerical artifact.
(The plots were taken from Ref.~\cite{janka.ref_kit03}.)
}
\label{janka.fig1}
\end{figure*}

%
%
%
%
%
%
%
%

\begin{figure*}[t!]
\tabcolsep=0.5mm
\begin{tabular}{lcr}
\epsfclipon \epsfig{file=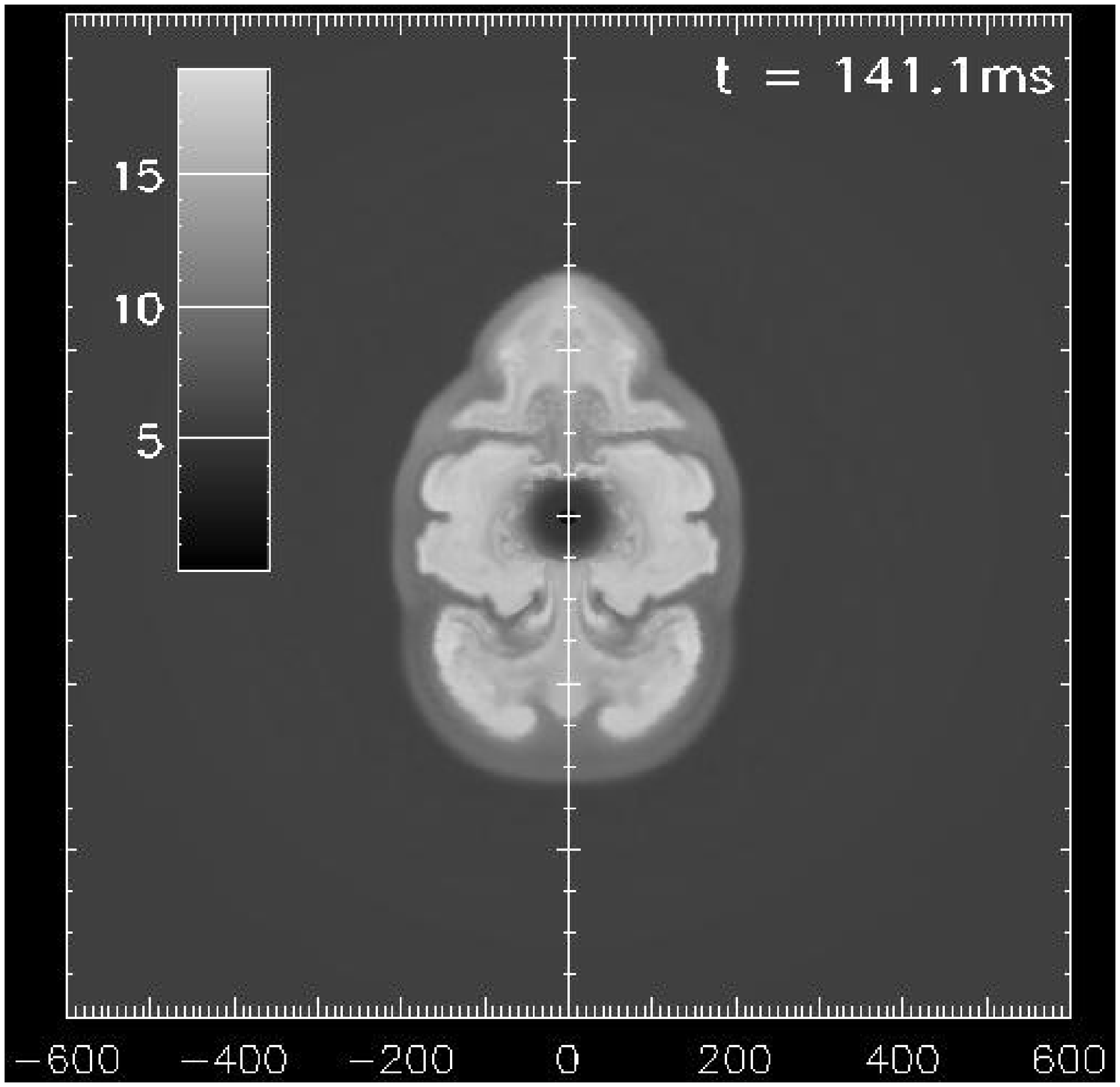,width=0.33\hsize} &
\epsfclipon \epsfig{file=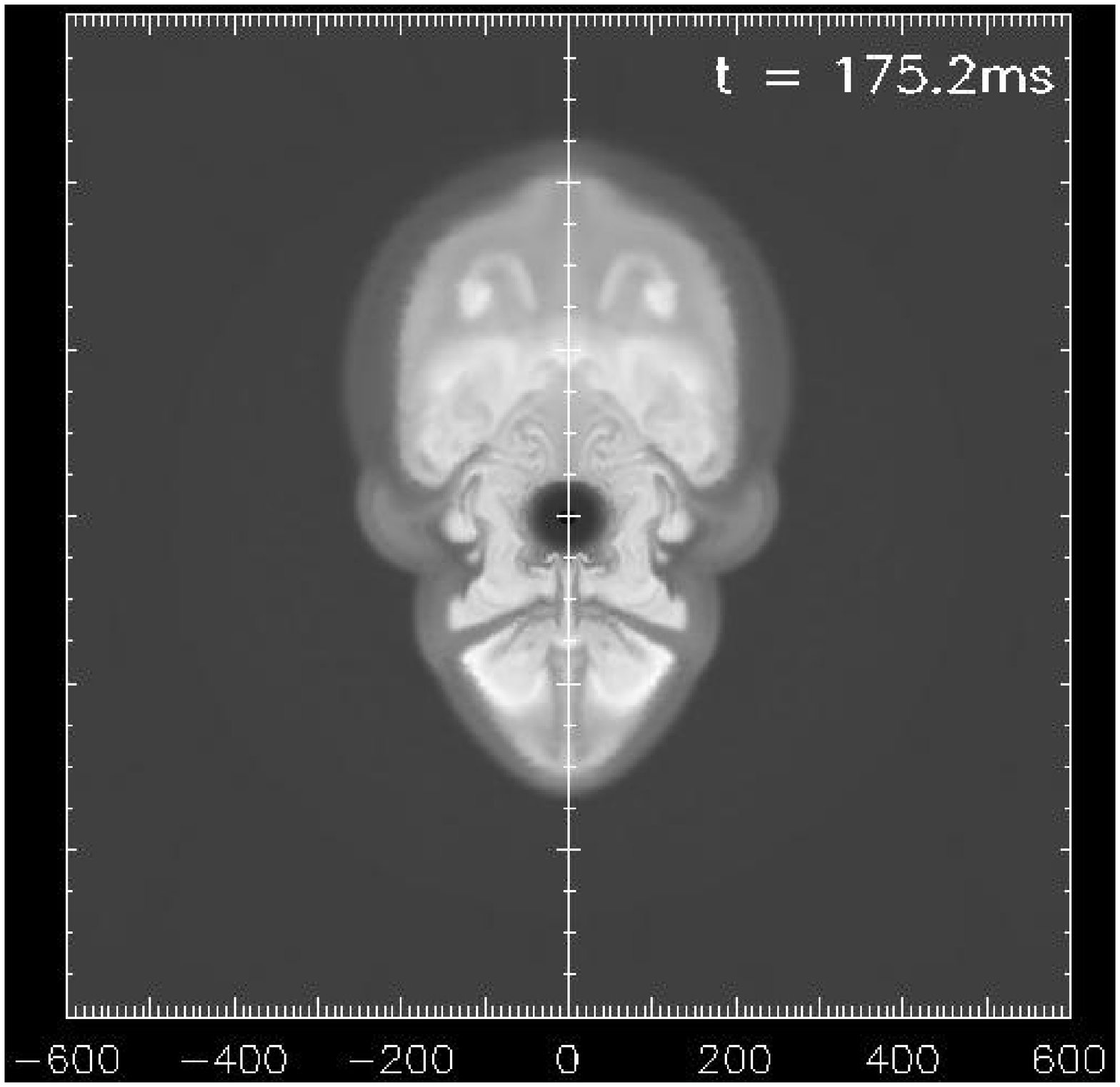,width=0.33\hsize} &
\epsfclipon \epsfig{file=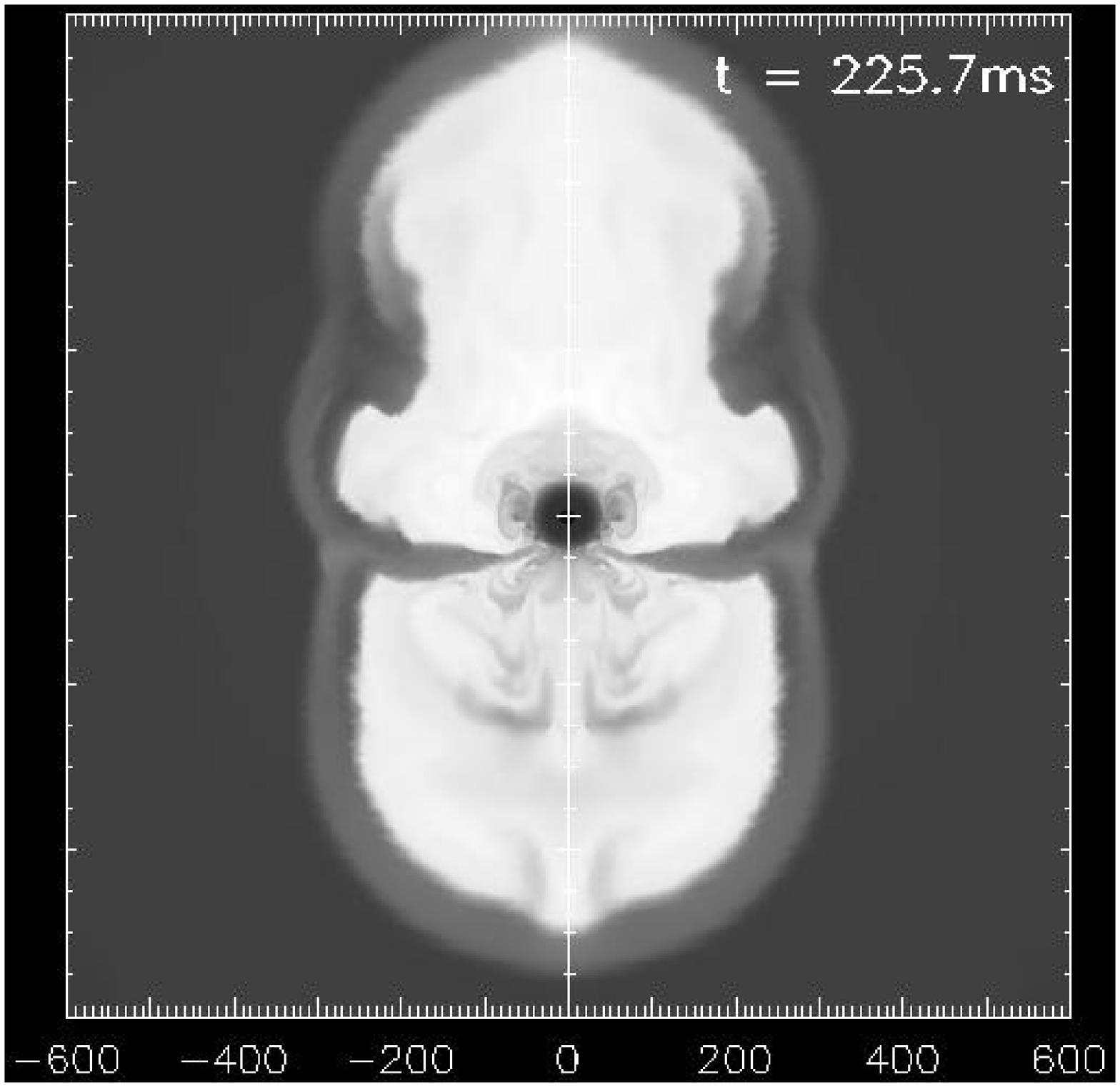,width=0.33\hsize}
\end{tabular}
\vspace{-25pt}
\caption{\small
Three stages (at postbounce times of 141.1$\,$ms,
175.2$\,$ms, 
and 225.7$\,$ms from left to right)
during the evolution of a (non-rotating) 11.2$\,$M$_{\odot}$
progenitor model from Ref.~\cite{janka.ref_wooetal02}, visualized
in terms of the entropy. The scale is in km. The dense NS
is visible as low-entropy circle at the center.
The computation was performed in
spherical coordinates, assuming axial symmetry, and employing
the ``ray-by-ray plus'' variable Eddington factor
technique~\cite{janka.ref_ramjan02}
for treating $\nu$ transport in multi-dimensional
SN simulations. Equatorial
symmetry is broken on large scales soon after bounce, and
low-mode convection begins to dominate the flow between
the NS and the strongly deformed SN shock.
The model continues to develop a probably weak explosion, the
energy of which was not determined before the simulation had
to be stopped because of CPU time limitations.
}
\label{janka.fig2}
\end{figure*}

\section{EXPLOSION MODELS}

Simulations leading to explosions can be reported for stars 
near the low-mass end of SN progenitors, namely for
stars in the mass range of $\sim$8--10$\,M_{\odot}$ with 
O-Ne-Mg cores~\cite{janka.ref_nom87}, and for an 11.2$\,M_{\odot}$
progenitor~\cite{janka.ref_wooetal02}, both characterized
by small cores of less than $\la$1.3$\,M_{\odot}$ (O-Ne-Mg
or Fe, respectively) and a steep density decrease (entropy
increase) outside.

\subsection{Stars in the 8--10$\,M_{\odot}$ range with O-Ne-Mg cores}

The main improvement of our new simulations of O-Ne-Mg core collapse
--- which we did so far only in spherical symmetry ---
compared to previous approaches is the more accurate, 
spectral treatment of $\nu$ transport and $\nu$-matter
interactions. Using the nuclear equation of state (EoS) of
Lattimer \& Swesty~\cite{janka.ref_latswe91} and more recently 
also that of Hillebrandt \& Wolff~\cite{janka.ref_hiletal84}, we
could {\em not} confirm the prompt explosions found in calculations
with simpler $\nu$ treatment~\cite{janka.ref_hiletal84}.
The shock is created much deeper inside the collapsing core
than in the ``old'' models (cf.\ also 
Ref.~\cite{janka.ref_baretal87}), typically at a mass coordinate
around $\sim$0.45$\,M_{\odot}$~\cite{janka.ref_kit03}
(defined by the moment when the postshock entropy first exceeds
3$\,k_{\mathrm B}$ per nucleon), and it stalls
(defined by the time when the postshock velocity becomes negative)
only 1.2$\,$ms later at $\sim$0.8$\,M_{\odot}$, still
well inside the neutrinosphere and {\em before} energy losses by
the prompt $\nu_e$ burst could have contributed to its 
damping~\cite{janka.ref_kit03}.
We also do not find the powerful shock revival by $\nu$ heating
as seen in Ref.~\cite{janka.ref_maywil88} and for some choice of
the nuclear EoS by Fryer et al.~\cite{janka.ref_fry99}.
Instead, the shock
continuously expands due to the monotonically decreasing preshock
density and the steep density decline at the
interface between C-O shell and He shell (Fig.~\ref{janka.fig1}).
At the end of
our simulation the mass accretion rate by the shock has
correspondingly dropped to less than 0.03$\,M_{\odot}\,$s$^{-1}$.
Although our simulation is not yet finally conclusive in this
point, we see indications that a $\nu$-driven wind begins
to fill the volume between SN surface and shock and   
will lead to mass ejection with a rather low explosion energy
(a few $10^{50}\,$erg due to the wind power and nuclear burning).
Little nickel production ($\sim$0.01$\,M_{\odot}$),
can be expected, corresponding to the wind mass loss rate that
can be estimated for an initial $\nu_e+\bar\nu_e$ luminosity of 
$\sim 8\times 10^{52}\,$erg$\,$s$^{-1}$
(Fig.~\ref{janka.fig1}) using the equations in
Refs.~\cite{janka.ref_nuwind}.
The baryonic mass of the NS will be very close 
to or only some 0.01$\,M_{\odot}$
less than the C+O core mass (which is 1.38$\,M_{\odot}$).
These findings are very similar to the outcome
of simulations of accretion induced white dwarf collapse to
NSs (AICs)~\cite{janka.ref_woobar92}. 

A weak explosion of an 8--10$\,M_{\odot}$ star has been
suggested as an explanation of the observed properties of the
Crab SN remnant~\cite{janka.ref_nom82}. The wind-driven
explosion seen in our models, however, does not provide the
conditions for the low-entropy, low-$Y_e$ r-process nucleosynthesis
discussed for prompt explosions of collapsing O-Ne-Mg cores
in previous work~\cite{janka.ref_promptrp}.
R-processing in the high-entropy environment of the 
$\nu$-driven baryonic wind can also not be expected
to take place, because sufficiently high entropies, 
short expansion timescales, and low proton-to-baryon ratios
require the NS to be very massive 
($\sim$2$\,M_{\odot}$) and compact ($\la$10$\,$km) (e.g.,
Refs.~\cite{janka.ref_nuwind}).
It is therefore unclear how O-Ne-Mg core collapse events could
contribute to the production of high-mass r-process nuclei.

\begin{figure}[htb]
\begin{minipage}[t]{80mm}
\epsfclipon \epsfig{file=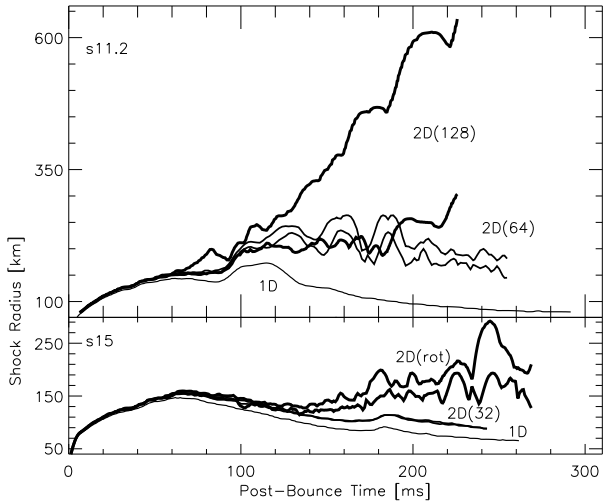,width=\hsize}  
\end{minipage}
\hspace{\fill}
\begin{minipage}[t]{75mm}
\vspace{-7.6truecm}
\caption{\small 
Maximum and minimum radii of the shock (giving a rough
measure of deformation)
vs.\ time for 2D simulations with a polar coordinate grid
(models ``2D''),
compared to 1D results with the same physics, but no 
convection (models ``1D'').
{\em Top:} 11.2$\,M_{\odot}$ star~\cite{janka.ref_wooetal02}, 
simulated with a full
180$^{\mathrm o}$ grid (Model 2D(128))
and with a 90$^{\mathrm o}$ angular wedge around the
equatorial plane (Model 2D(64)). 
{\em Bottom:} 15$\,M_{\odot}$ star
(Model~s15s7b2 of Ref.~\cite{janka.ref_woowea95})
without (Model 2D(32)) and with rotation (Model 2D(rot); see
also~\cite{janka.ref_buras03,janka.ref_mueller04}).
In the former 2D simulation a
90$^{\mathrm o}$ wedge around the equator was used, in the
latter a 90$^{\mathrm o}$ grid from pole to equator.
}
\label{janka.fig3}
\end{minipage}
\end{figure}

\subsection{Massive stars with iron core}

Progenitor stars in the mass range between 11$\,M_{\odot}$
and 25$\,M_{\odot}$ were found to neither explode by the 
prompt bounce-shock mechanism nor by the delayed 
$\nu$-heating mechanism in spherically symmetric 
simulations~\cite{janka.ref_jan04c}, in agreement with 
results of other groups~\cite{janka.ref_BTsims}.
This implies that state-of-the-art SN models do not 
support suggestions that r-process nucleosynthesis might
occur in prompt explosions of low-mass ($\sim$11$\,M_{\odot}$)
progenitors~\cite{janka.ref_sumi01}.

We have also started to perform core-collapse simulations 
for the mentioned stellar mass range in 2D, using a polar
coordinate grid and assuming azimuthal 
symmetry~\cite{janka.ref_buras03,janka.ref_jan04a}. Because of
the energy-dependent treatment of $\nu$ transport and 
$\nu$-matter interactions the requirements of CPU
time are substantial, and we were so far able to perform
only a handful of such 2D calculations. To limit the need of
computer resources we initially constrained the computational
volume to a $\sim$90$^{\mathrm{o}}$ wedge (between roughly
$+$45$^{\mathrm{o}}$ and $-$45$^{\mathrm{o}}$ around the 
equatorial plane and periodic boundary conditions for 
nonrotating models and between 0$^{\mathrm{o}}$ and 
90$^{\mathrm{o}}$ with reflecting boundaries for rotating ones),
using an angular resolution of $\sim$1.4$^{\mathrm{o}}$. 
This choice was also motivated by 2D models of the
first generation which were able to obtain explosions with
a similar setup due to the help of hot-bubble 
convection~\cite{janka.ref_SN2D}.

First simulations of stars of 11.2, 15, and 20$\,M_{\odot}$
with a 90$^{\mathrm{o}}$ wedge, however, did not produce 
explosions~\cite{janka.ref_buras03}. 
This suggests that the success of the previous calculations
with simplified transport was most probably connected with
the use of transport approximations. One of our new models,
a 15$\,M_{\odot}$ star (Model s15s7b2 of 
Ref.~\cite{janka.ref_woowea95}), was also collapsed with a
modest rotation: The pre-collapse core was assumed
to rotate (essentially rigidly) with a period of 12$\,$s,
i.e. $\Omega \approx 0.5\,$rad$\,$s$^{-1}$ 
(see Fig.~1 in \cite{janka.ref_mueller04}), which is 
significantly faster --- but not orders of magnitude 
faster --- than predicted by the latest stellar evolution
models~\cite{janka.ref_heger04}. Even this
``modest'' amount of rotation turned out to make a big 
difference. As visible in the lower panel of 
Fig.~\ref{janka.fig3}, the shock expands to a much larger
radius than in the 2D simulation without rotation
(although both simulations did not lead to explosions 
within $\sim$270$\,$ms of postbounce evolution), mostly
because of the influence of centrifugal forces and the
more violent convection in a more extended region of 
$\nu$ heating~\cite{janka.ref_buras03}.

\subsection{Nonradial shock instabilities and 
low-mode postshock convection}

While the 11.2$\,M_{\odot}$ simulation with 90$^{\mathrm{o}}$ 
wedge did not explode (see Fig.~\ref{janka.fig3}, upper panel),
the same model with a full 180$^{\mathrm{o}}$
grid developed a presumably weak explosion
(Figs.~\ref{janka.fig2} and \ref{janka.fig3}). Equatorial
symmetry is broken on large scales some 10$\,$ms after 
convective activity in the
postshock region has started (at $\sim$50$\,$ms p.b.), and
low modes ($l = 1,2$) begin to dominate the flow pattern between
NS and shock after about 140$\,$ms p.b., a phenomenon which 
might be linked to the observed large recoil velocities of young
pulsars~\cite{janka.ref_pulsarkicks}.
The convection becomes significantly more 
violent than in the 90$^{\mathrm{o}}$ simulation where obviously
important degrees of freedom were suppressed.
This latter fact was emphasized in an interesting paper 
by Blondin et al.~\cite{janka.ref_blondin04}, who found a similar
development of low-mode asymmetries in the postshock flow and
shock oscillations as a consequence of the instability 
of standing accretion shocks (``SASI'') against non-radial 
perturbations that are amplified in a ``vortical-acoustic
feedback cycle''~\cite{janka.ref_foglizzo02}.

Blondin et al.\ considered idealized conditions in their
numerical studies, assuming steady-state mass accretion 
with a constant rate, a fixed inner boundary radius, a simple
ideal-gas EoS, and ignoring $\nu$ heating and
cooling. We decided to test the effects of $\nu$'s and of
the size of the angular wedge (and thus of the possible modes)
in a separate set of simulations with a realistic EoS
but an approximative treatment of the $\nu$
transport (as described in Refs.~\cite{janka.ref_pulsarkicks}),
which allowed us to save computer time and thus to perform faster
calculations with higher resolution, and to run more models.
The mass accretion rate was given by the collapse of a 
15$\,M_{\odot}$ progenitor star and the NS was replaced
by a gravitating point mass inside 
a contracting inner boundary that followed the behavior of the
shrinking NS in our full-scale SN simulations with
detailed transport physics.

With this setup we confirmed that 2D simulations with a 
180$^{\mathrm o}$ grid can yield explosions even if models with
90$^{\mathrm o}$ wedge do not explode. In these 
studies, with $\nu$ effects switched on or off, we found
that $\nu$ losses promote the action of the vortical-acoustic
cycle by allowing matter to settle on the proto-NS
surface. 
Our results show that corresponding shock deformation produces
growing perturbations in the postshock flow which 
can accelerate the onset of convective overturn even in cases where
the infall timescale is initially shorter than the growth timescale
of Ledoux convection. We also
see that dipolar shock oscillations become more violent in case
of a rapid contraction of our inner boundary (mimicing a NS
that becomes rapidly more compact due to a soft nuclear 
EoS), thus releasing gravitational
binding energy which partly is converted to turbulent kinetic 
energy of the postshock flow. In this case $l=1,2$ modes begin to 
dominate faster and the SASI conspires with 
convective instability to establish favorable conditions for 
a high efficiency of $\nu$ heating behind the shock. 
Both nonradial instabilities differ characteristically in the way
how anisotropies emerge: Convective overturn is driven by the 
negative entropy gradient in the $\nu$-heated layer, which
becomes Rayleigh-Taylor (RT)
unstable first on small scales, while the SASI starts from 
vorticity producing sound wave interactions with the shock and
induces dipolar oscillations of the postshock layer 
before convective activity is initiated and the typical RT 
mushrooms become visible.

\begin{figure*}[t!]
\tabcolsep=0.5mm
\begin{tabular}{lcr}
\epsfclipon \epsfig{file=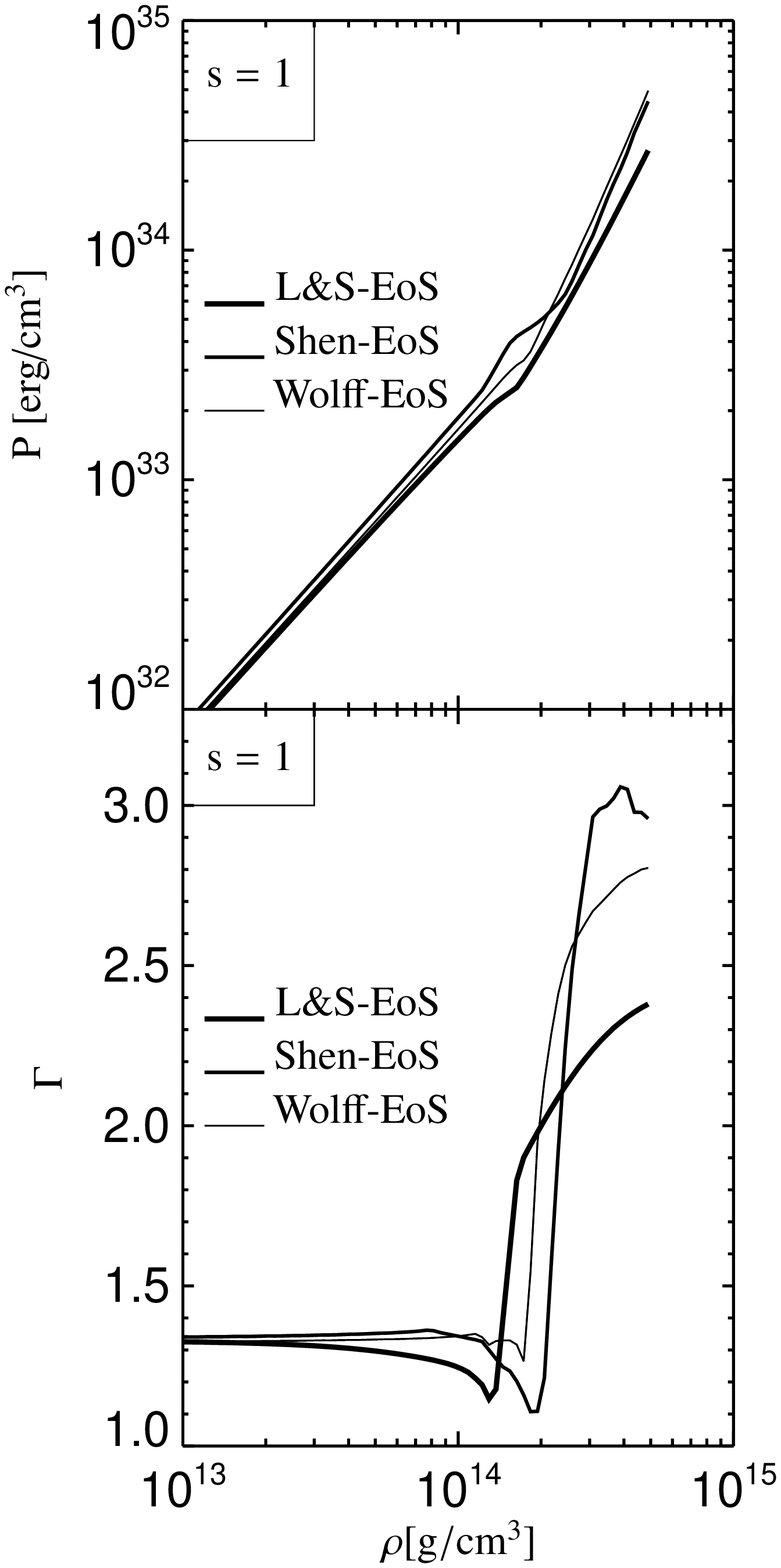,width=0.30\hsize} &
\epsfclipon \epsfig{file=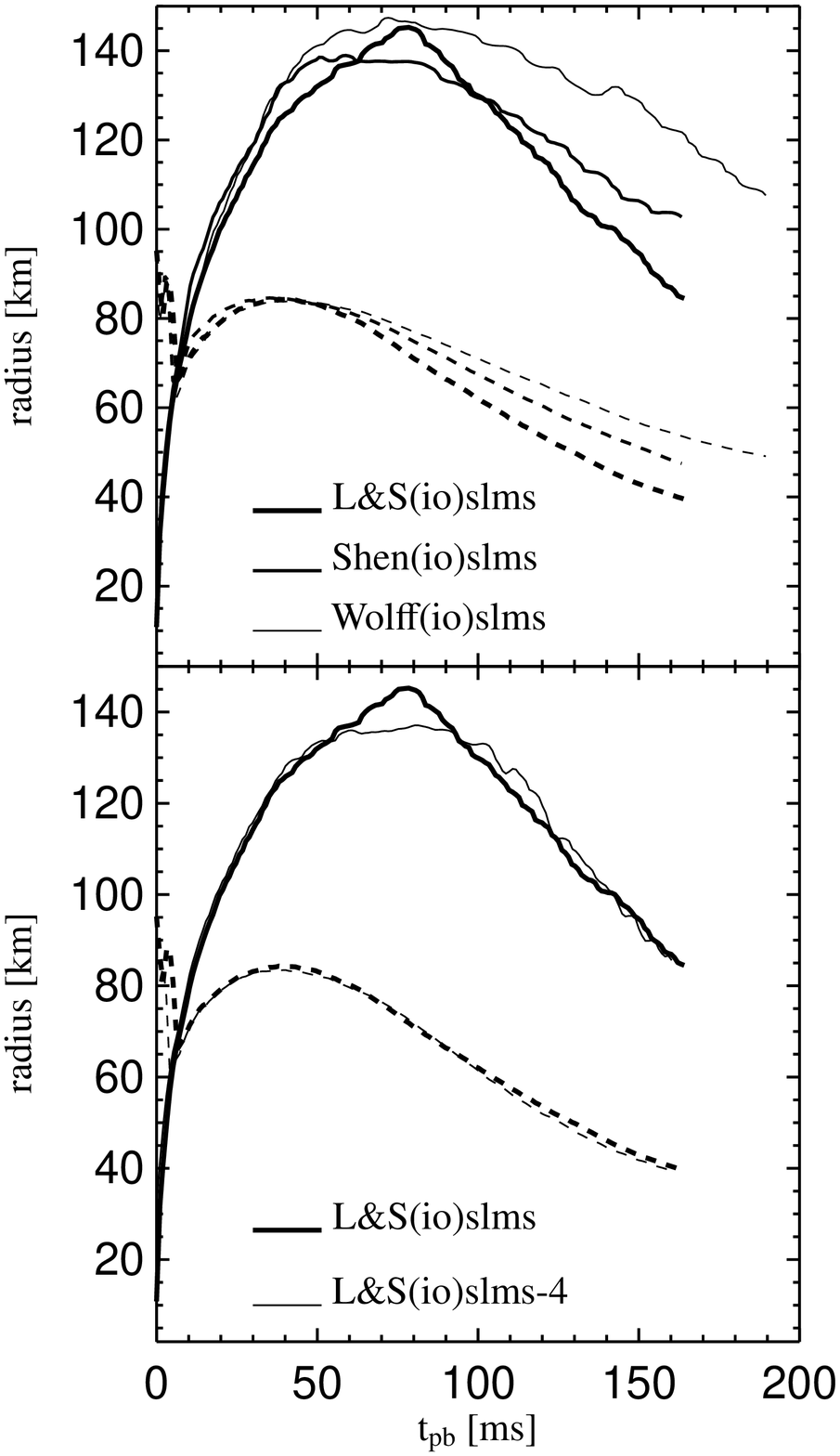,width=0.348\hsize} &
\epsfclipon \epsfig{file=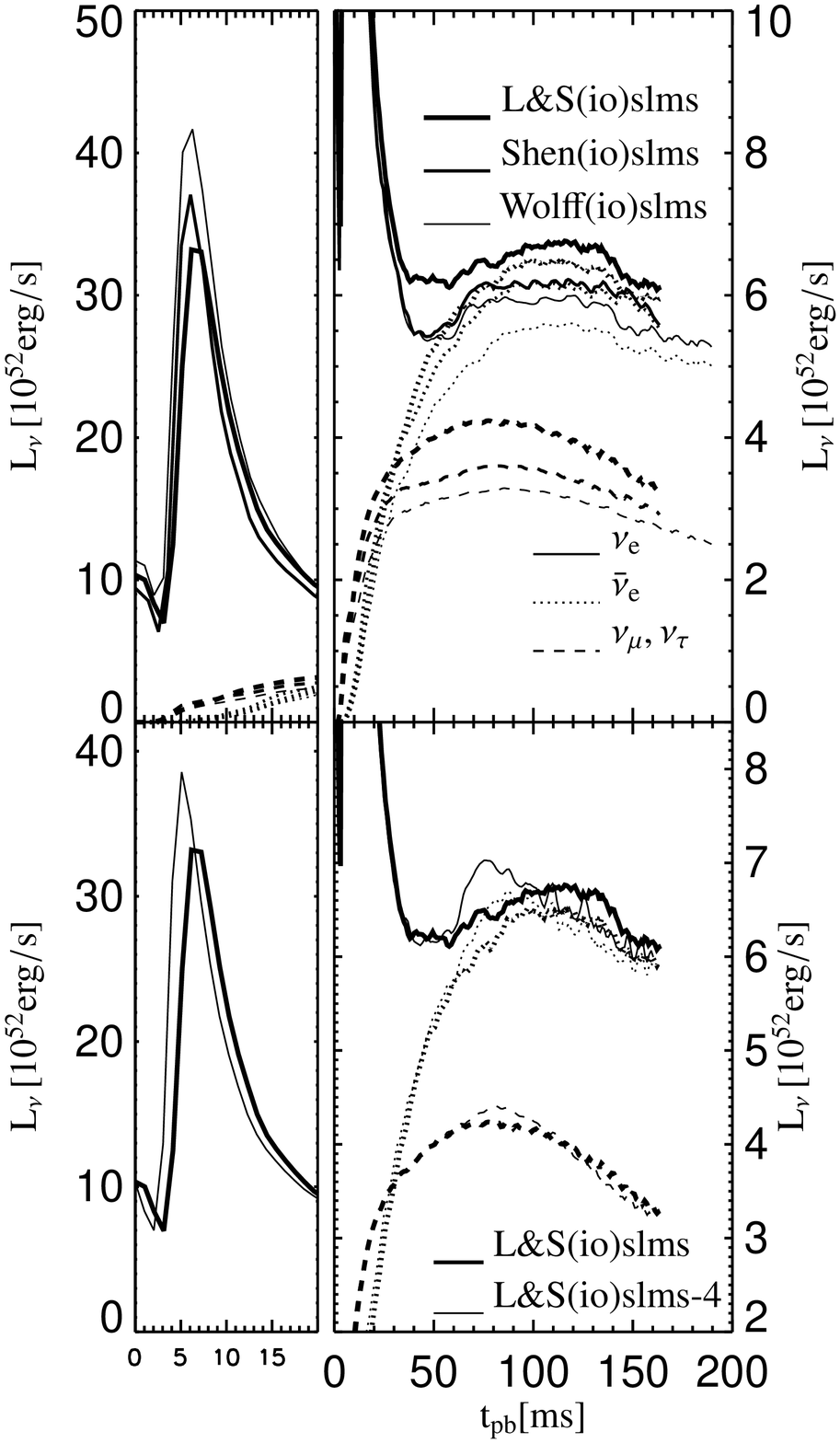,width=0.352\hsize}
\end{tabular}
\vspace{-25pt}
\caption{\small
{\em Left:} Pressure (top) and adiabatic index
$\Gamma \equiv \left ({\mathrm d}\ln P/{\mathrm d}\ln\rho\right )_s$  
(bottom) vs.\ mass density
for an entropy $s = 1\,k_{\mathrm B}$ per nucleon (and
$Y_e = 0.4$) for the EoSs
of Ref.~\cite{janka.ref_hiletal84} (``Wolff'', thin lines),
Ref.~\cite{janka.ref_shenetal98} (``Shen'', medium)
and Ref.~\cite{janka.ref_latswe91} (``L\&S'', thick).
{\em Middle:} Shock positions (solid lines) and 
neutrinospheric radii of $\nu_e$ (dashed)
as functions of time for collapse simulations of a 15$\,M_{\odot}$
progenitor (Model~s15a28~\cite{janka.ref_wooetal02})
with the three nuclear EoSs (top) and with two
different EoSs at densities $\rho < 10^{11}\,$g$\,$cm$^{-3}$ 
(L\&S compared to an ideal gas EoS of $e^-$, $e^+$, $\gamma$'s, 
and Boltzmann gases of $n$, $p$, $\alpha$ and a 
representative heavy nucleus in NSE; bottom).
{\em Right:} Prompt $\nu_e$ burst (left panel) and postbounce
luminosities of $\nu_e$ (solid lines), $\bar\nu_e$ (dotted) and
heavy-lepton $\nu$'s and $\bar\nu$'s (dashed) for the
simulations of the 15$\,M_{\odot}$ star with the three different
nuclear EoSs (top) and the two different low-density EoSs (bottom).
(The plots were taken from Ref.~\cite{janka.ref_marek03}.)
}
\label{janka.fig4}
\end{figure*}

\subsection{Nuclear physics sets the stage}

Nuclear physics therefore governs not only the
collapse phase, where electron captures on nuclei were
recently included in a much improved way and were found to determine
the position of shock formation and the structure of the layers
the shock expands into~\cite{janka.ref_ecapture}
(also W.R.~Hix and G.~Mart\'{\i}nez-Pinedo, this conference).
Nuclear physics and in particular the properties of the nuclear
EoS also determine the contraction and size of the nascent NS
and thus may influence the growth of nonradial
instabilities and anisotropies in the postshock accretion flow
during the $\nu$-heating phase.

In fact, differences in the nuclear EoS have interesting 
consequences for core collapse, bounce conditions, shock formation,
and postshock evolution
in 1D simulations~\cite{janka.ref_marek03}
(for a brief summary, see~\cite{janka.ref_jan04c}).
Figure~\ref{janka.fig4} shows selected results for three 
different EoSs (Lattimer \& Swesty~\cite{janka.ref_latswe91},
Shen et al.~\cite{janka.ref_shenetal98}, and Wolff \& 
Hillebrandt~\cite{janka.ref_hiletal84}). The softest of them
(L\&S) leads to the highest densities at bounce and the smallest
enclosed mass of the shock formation position. Lateron the
nascent NS contracts most rapidly, forcing the shock
to retreat much more quickly than in case of, in particular, the
stiff Wolff EoS (Fig.~\ref{janka.fig4}, upper middle panel). 
These differences
affect the $\nu$ luminosities during the $\nu_e$ burst and
the postbounce accretion phase (Fig.~\ref{janka.fig4}, 
upper right panel). While for the Wolff EoS the $\nu_e$ 
release in the burst is highest, 
a more compact NS and thus hotter neutrinosphere
causes higher post-burst $\nu$ luminosities and mean energies.
In the lower middle and right panels we show the result of a 
test which we performed with a four-nuclei NSE-EoS replacing 
the low-density ($\rho < 10^{11}\,$g$\,$cm$^{-3}$) part of the
L\&S EoS (where an error in the treatment of $\alpha$-particles
has recently been discovered; C.\ Fryer, J.\ Lattimer, personal
communication). The conditions in the postshock 
layer were hardly affected because the gas is disintegrated into
free nucleons at the high 
entropies encountered in simulations with relativistic gravity.
Minor differences in the evolution of the
shock radius (Fig.~\ref{janka.fig4}, lower middle panel) and
$\nu$ luminosities (Fig.~\ref{janka.fig4}, lower right panel)
were caused by differences in the mass accretion rate associated
with the EoS treatment in the infall region ahead of
the shock.

\section{SUMMARY}

Non-radial instability of the accretion shock can amplify vorticity
in the postshock flow and thus can support the growth of convection
in the $\nu$-heated layer. It may be an important ingredient 
for eventually robust explosions by the $\nu$-driven mechanism. 
Low modes in the flow can develop when the explosion sets in slowly.
The relative importance of the different instabilities seems to
depend on $\nu$ cooling and heating on the one hand and the
high-density EoS, which controls the contraction of the nascent 
NS, on the other. Simulations require the use of a
full 180$^{\mathrm{o}}$ grid and ultimately may have to be done
in 3D.

{\small
\medskip\noindent
{\bf Acknowledgements.}
We are grateful to K.~Nomoto, A.~Heger, S.~Woosley, and
M.~Limongi for providing us with their progenitor data.
Supercomputer time at the John von Neumann Institute for
Computing in J\"ulich and the Rechenzentrum Garching is
acknowledged.
}

\end{document}